\documentclass[12pt,preprint,dvips]{aastex}

\usepackage{graphicx}
\usepackage{dcolumn}
\usepackage{bm}

\newcounter{sub}
\newcounter{subeqn}[sub]
\renewcommand{\thesubeqn}{\alph{subeqn}}
\renewcommand{\theequation}{\thesub\thesubeqn}

\def\be{\begin{equation}}
\def\ee{\end{equation}}

\def\ms{\left[}
\def\rs{\right]}

\def\st{\stepcounter{sub}}

\def\bea{\begin{eqnarray}}
\def\eea{\end{eqnarray}}

\newcommand\xxi{{\mbox{\boldmath $\xi$}}}
\newcommand\xis{{\mbox{\scriptsize{$\xxi$}}}}
\newcommand\mmu{{\mbox{\boldmath $\mu$}}}
\newcommand\llambda{{\mbox{\boldmath $\lambda$}}}

\newcommand\eeta{{\mbox{\boldmath $\eta$}}}
\newcommand\etas{{\mbox{\scriptsize{$\eeta$}}}}
\newcommand\nab{\mbox{\boldmath $\nabla$}}

\newcommand\psip{{\mbox{${\psi^+}$}}}
\newcommand\chip{{\mbox{${\chi^+}$}}}
\newcommand\phip{{\mbox{${\phi^+}$}}}

\def\k{{\bf k}}
\def\l{{\bf l}}
\def\m{{\bf m}}

\def\r{{\bf r}}

\newcommand\A{{\bf A}}
\newcommand\I{{\bf I}}

\newcommand\no{\nonumber}

\newcommand\GTP{{\rm GTP}}
\newcommand\GDP{{\rm GDP}}
\newcommand\Tu{{\rm T}}
\newcommand\MT{{\rm MT}}

\begin{document}


\title{From a quantum mechanical description of the assembly processes in microtubules to their semiclassical
nonlinear dynamics }

\author{Vahid Rezania$^{1, 2, 3}$ and Jack Tuszynski$^1$}

\affil{1- Division of Experimental Oncology, Cross Cancer Institute\\
11560 University Avenue, Edmonton, AB T6G 1Z2
Canada}
\affil{2- Department of Science, City Center Campus, \\Grant MacEwan College, Edmonton, AB T5J 2P2, Canada}
\affil{3- Institute for Advanced Studies in Basic Sciences,
          Zanjan 45195, Iran}




\begin{abstract}
In this paper a quantum mechanical description of the
assembly/disassembly process for microtubules is proposed. We
introduce creation and annihilation operators that raise or lower
the microtubule length by a tubulin layer.   Following that, the
Hamiltonian and corresponding equations of motion are derived that
describe the dynamics of microtubules.   These Heisenberg-type equations are then
transformed to semi-classical equations using the method of coherent
structures. The latter equations are very similar to the
phenomenological equations that describe dynamic instability of
microtubules in a tubulin solution.
\end{abstract}

\maketitle

\section{Introduction}\label{int}
\noindent In most multicellular organisms, the interior of each cell
is spanned by a dynamic network of molecular fibers called the
\emph{cytoskeleton} (`skeleton of the cell'). The cytoskeleton gives
a cell its shape, acts as a conveyor for molecular transport, and
organizes the segregation of chromosomes during cell division,
amongst many other activities. The complexity and specificity of its
functions has given rise to the theory that along with its
structural and mechanical roles, the cytoskeleton also acts as an information
processor \citep{Alb85}, or simply put the ``cell's nervous system"
\citep{Ham87}. Microtubules are the cytoskeleton's most studied
components, and over the years many models of microtubular
information processing have been proposed.  A microtubule is a
hollow cylinder, a rolled-up hexagonal array of tubulin dimers
arranged in chains along the cylinder
(`protofilaments'). Within cells, microtubules come in bundles held
together by `microtubule associated proteins' (MAPs). The geometry,
behavior and exact constitution of microtubules varies between cells
and between species, but an especially stable form of microtubule
runs down the interior of the axons of human neurons. Conventional neuroscience
at present ascribes no computational role to them, but models exist
in which they interact with the membrane's action potential
\citep{BT97,Pri06}.

Microtubules are very dynamic bio-polymers that simply lengthen and/or shorten
repeatedly at the macroscopic level during a course of time.
At the microscopic level, however, several biochemical reactions
are taking place in order for an individual microtubule to undergoe an assembly or
disassembly process.
This dynamical behavior of microtubules (so-called dynamic instability)
has attracted many investigators for decades to examine microtubules'
behavior in many aspects.  See section 2 for further details.

Though there is no systematic description for the microtubule's assembly/disassembly
process at the microscopic level, several theoretical models are proposed to
describe the macroscopic lengthening/shortening of microtubules using nonlinear
classical equations \citep{DL93,Dog95,Bic97,DY98,BR99}.
In spite of their agreement with experimental results, the latter studies are more or
less phenomenological.  Therefore, several features of the
microtubule's assembly/disassembly process might not be captured.

In this paper, we propose a systematic model for the microtubule's
assembly/disassembly process at the microscopic level using a
first-principles quantum mechanical approach. In this model we consider an individual
microtubule with length $L$ consisting of $N$ tubulin layers viewed here as a quantum state $|N\rangle$.
The state can be raised/lowered by creation/annihilation operator
(i.e. polymerization/depolymerization process) to $|N+1\rangle/|N-1\rangle$
state.  The corresponding microtubule is then longer/shorter by one tubulin
layer from the original one.
Based on the chemical binding reactions that are taking place during microtubule
polymerization, a quantum mechanical Hamiltonian for the system is proposed.
Equations of motion are then derived and transformed from the purely quantum mechanical
description to a semi-classical picture using the method of coherent
structures.  The resulting nonlinear field dynamics
is richer than the previous phenomenological descriptions and
includes both localized energy transfer and oscillatory solutions.

\section{Microtubule assembly background}
\noindent
A very rigid and typically several micrometers long rod-like polymer plays an
essential role during cell division.    The so-called
microtubule (MT) is assembled by tubulin polymerization
in a helical lattice.  These protein polymers are
responsible for several fundamental cellular processes, such as
locomotion, morphogenesis, and reproduction \citep{ALRW94}.
It is also suggested that MTs are responsible for transferring energy
across the cell, with little or no dissipation.

Both \emph{in vivo} and \emph{in vitro} observations
confirmed that an individual microtubule switches stochastically
between assembling and disassembling states that makes MTs
highly dynamic structures \citep{MK84a,MK84b}.
This behavior of MTs is referred to as dynamic instability.
Dynamic instability of microtubules is a nonequilibrium process that has
been subject of extensive research for the past two decades.
It is generally believed that the instability starts from the hydrolysis
of guanosine triphosphate (GTP) tubulin that follows by converting GTP
to guanosine diphosphate (GDP).   This reaction is exothermic and
releases $\sim 8 kT$ energy per reaction \citep{WIS89}, i.e. approximately
$0.22$ eV per molecule \citep{EAHH89}.  Here $k$ is the Boltzmann's
constant and $T$ is the temperature.   Since GDP-bound tubulin favors
dissociation, an MT enters the depolymerization phase as the advancing
hydrolysis reaches the growing end of an MT.   This phase transition is
called a catastrophe.  As a result of this transition, MTs start breaking
down, releasing the GDP-tubulin in the solution.  In the solution,
however, reverse hydrolysis takes place and polymerization phase of
MTs begins.  The latter phase transition which comes after a catastrophe
is called a rescue.   Therefore, MTs constantly fluctuate between growth
and shrinkage phases.

Interestingly, \cite{OCB95} studied experimentally and theoretically MTs' assembly
to extract their catastrophe kinetics.
They proposed that a growing MT
may remember its past phase states by assessing growth of
both plus and minus ends of several individual MTs.   Their results
showed that while the minus end growth time follows an
exponential distribution, the plus end fits a gamma distribution.
The exponential (gamma) distribution suggests a
first (non-first) order transition between growing and shrinking
phases.  Statistically, the exponential distribution
represents that the new state happens independently of the
previous state.  As a result, an MT with first order catastrophe
kinetics does not remember for how long it has been growing.
  In contrast, the
catastrophe frequency of an MT with non-first order kinetics would
depend on its growth phase period.  The gamma distribution suggests
that the catastrophe frequency is close to zero at early times,
increases over time and reaches asymptotically a plateau.
This is consistent with observations that the
catastrophe events are more likely at longer times.
\cite{OCB95} concluded that such behavior implies that a
`crude form of memory' may be built in MT's dynamic instability.
As a result, a microtubule would go through an `intermediate state'
before a catastrophe event takes place.

The dynamics of transitions between growing and shrinking states is still
a subject of controversy.  It is suggested that a growing MT has a
stabilizing cap of GTP tubulin at the end which keeps it from
disassembling \citep{MK84a,MK84b}.  Whenever MT loses its cap,
it will undergo the shrinking state.   Several theoretical and
experimental studies have been devoted to the cap model.  For the purpose of this paper,
we emphasize the link between GTP hydrolysis and the switching process
from growing to shrinking of an MT.   GTP hydrolysis is a subtle biochemical
process that carries a quantum of a biological energy and thus allows us
to make a link between quantum mechanics and polymer dynamics.
We return to this theme later in the paper but first discuss the
statistical methods used in this area.

\subsection{Ensemble dynamics of microtubules}
\noindent
As we discussed earlier the MT dynamical instability has been the subject of
numerous studies. Although the dynamical instability of MTs is a
nonlinear and stochastic process,
investigators modeled their averaged behaviors using a simple model.  Introducing
$p_g(x,t)$ and $p_s(x,t)$ as the probability density of a growing and shrinking tip,
respectively, of an MT with length $x$ at time $t$, \cite{DL93} proposed the following
equations for the time evolution of an individual MT:
\st
\bea \label{grow0}
&&\partial_t p_g = - f_{gs} p_g + f_{sg} p_s - v_g \partial_x p_g,\\
\st \label{shrink0}
&&\partial_t p_s =  f_{gs} p_g - f_{sg} p_s - v_s \partial_x p_s.
\eea
Here $f_{gs}$ and $f_{sg}$ are the transition rates from a growing to
a shrinking state and vice versa.  The average speeds of the MT in the assembly and
disassembly states are given by $v_g$ and $v_s$, respectively.  See also
\cite{Bic97}, \cite{DY98} and \cite{BR99}.

Random fluctuations about the MT's tip location can be also modeled by adding a
diffusive term in the above equations:
\st
\bea \label{grow}
&&\partial_t p_g = - f_{gs} p_g + f_{sg} p_s - v_g \partial_x p_g + D_g\partial_{xx}p_g,\\
\st \label{shrink}
&&\partial_t p_s =  f_{gs} p_g - f_{sg} p_s - v_s \partial_x p_s + D_s\partial_{xx}p_s,
\eea
where $D_g$ and $D_s$ are the effective diffusion constants in the two states \citep{FHL94,FHL96}.

Equations (\ref{grow}) and (\ref{shrink}) describe the overall dynamics of an individual MT
without considering the dynamics of GDP and GTP tubulin present in the solution.  It is
clear that the MTs are growing faster in the area with a higher concentration of GTP tubulin.
Using this fact, \cite{Dog95} generalized the above model by incorporating the tubulin dynamics.  They added
two more equations to the above system:
\st
\bea \label{c_T}
&&\partial_t c_T = - v_g s_0  p_g + k c_D + D\nabla^2 c_T,\\
\st \label{c_D}
&&\partial_t c_D =  v_s s_0  p_s - k c_D + D\nabla^2 c_D,
\eea
where $c_T$ and $c_D$ are average concentrations of GTP and GDP
tubulin, respectively.
$D$ is the diffusion coefficient, $k$ is the rate constant and
$0 \leq s_0\leq 1$.   In view of the link to quantum transitions between
GDP and GTP at the root of this problem we now introduce a method that allows a smooth
transition from quantum to classical (nonlinear) dynamics of MT assembly/disassembly
process.

\subsection{Method of Coherent Structures }

The method we use here is called the Method of Coherent Structures (MCS)
which has been developed in a number of papers and articles
\citep{TD89a,TD89b,TD89c,TD89d,DT90a,DT90b,Tus90} and is essentially
semiclassical in nature. The treatment is quantitative in that important
terms which are retained are calculated exactly and those which are very
small but nevertheless significant are discussed at a later stage and
their effect estimated.  The motivation for the method and a derivation
of the dynamical field equation are presented by \cite{TD89a} and a
discussion of the types of classical field solutions is presented by
\cite{TD89b}.  A fuller version has been published in the review paper
by \cite{Tus90} whereas a very brief overview is given by \cite{TD89c}.
It has been successfully applied to the phenomenon of superconductivity
\citep{TD89d,DT90a} and when combined with topological arguments
yields, for example, the correct temperature dependence of the critical
current density in low temperature superconductors.  One can also obtain
from MCS the position of phase boundaries in metamagnets where previously
only elaborate numerical techniques could provide this information
\citep{DT90b}.  Spatial correlations are fully incorporated using a
renormalization technique and quantum fluctuations have been
included also.
It has been demonstrated that even when the method is generalized
to include spin-dependent fields, the equation of motion for the field
is of the same form \citep{DT91} and the classical field equation is
also of the same form for both Boson and Fermion particles.  This does
not mean that the Fermionic character of the electrons disappears
because the statistics of the particles reappear in the
choice of the classical field which satisfies the physical boundary
conditions on the charge density.  The method is basically
non-relativistic although it could be readily generalized but here
we use the non-relativistic version.  

The starting point in the MCS is to write a generic form of
second-quantized Hamiltonian using one particle state
annihilation and creation operators:
\st
\be
H=\sum_\k\hbar\omega_\k q_\k^\dagger q_\k
+ \sum_{\k,~\l,~\m} \hbar\Delta_{\k,~\l,~\m}q_\k^\dagger q_\l^\dagger
q_\m q_{\k+\l-\m},
\ee
where the vectors $\k,~\l$ and $\m$ are shorthand labels for
quantum numbers of a complete orthonormal set of particle functions
in the usual way and we use the linear momentum conserving form for the
two-body interaction.  Depending on the system studied,
using Fermi-Dirac or Bose-Einstein statistics one can derive the
Heisenberg's equation of motion:
\st
\be\label{mm}
i\hbar\partial_t q_\k(\r,t)=[H,q_\k(\r,t)].
\ee
Now both sides of Eq. (\ref{mm}) are multiplied by
$\Omega^{-1/2} \exp(-i \eeta\cdot\r) a_\etas (t)$ and summed over $\eeta$.
At the same time the matrix elements $\omega_\k$ and $\Delta_{\k,~\l,~\m}$
are each expanded to second order in the deviations from the
point ($\k_0,~\l_0,~\m_0$).
After a considerable amount of algebra and a series of
transformations we find
\st
\bea\label{mm1}
i\partial_t\psi&=&\mu_0\psi + i \mmu_1\cdot\nab \psi
- \frac{1}{2} \sum_{i,j}(\mmu_2)_{ij}\partial^2_{x_ix_j}\psi\no\\
&&+~ \mu_3 \psi^+ \psi\psi  + i \mmu_4 \cdot \psi^+ \psi \nab\psi
+ i \mmu_5 \cdot \psi^+ (\nab \psi)\psi
+ i \mmu_6 \cdot (\nab \psi^+ )\psi\psi \no\\
&&+~ {\rm higher~ order~ terms},
\eea
where
\st
\be
\psi(\r,t) = \Omega^{-1/2} \sum_\etas \exp(-i \eeta\cdot\r)
a_\etas (t).
\ee
Here $\mu_i$ or $\mmu_i$ are constant parameters, determined by
matric elements $\omega_\k$ and
$\Delta_{\k,~\l,~\m}$ and their derivatives calculated at point
($\k_0,~\l_0,~\m_0$).
To convert Eq. (\ref{mm1}) to a PDE in a C-number field,
rather than an operator,
in MCS the center of expansion ($\k_0,~\l_0,~\m_0$) is selected to be a
critical or fixed point of the system.
The reason for this is that close to a critical point it is an
excellent approximation to replace the full quantum field, $\psi(r,t)$,
by a classical component, $\psi_c$ \citep{Ma76,Jac77,Ami78}:
\st
\be
\psi(\r,t)=\psi_c(\r,t) \hat{\I} + \hat{\phi}(\r,t),
\ee
where $\hat{\I}$ is the unit operator in Fock space,
$\psi_c$ is a c-number field, $\hat{\phi}$ is a quantum
mechanical operator with magnitude about
$|\hat{\phi}|\sim \hbar |\psi_c|$ \citep{DT95}.
See \cite{TSB97} for details.

In the next section we apply the MCS method to study the dynamic instability
of an individual microtubule.

\section{A quantum mechanical picture of the microtubule assembly processes }
\subsection{Particle states}
\noindent
For simplicity, we consider that MT polymerization to be a 1D process.
Consider an individual microtubule in a free tubulin solution
containing a large number of GTP-tubulin, GDP-tubulin and a pool of free GTP
molecules.  In this solution several processes take place (as well as their reverse reactions):\\
\noindent
(i) creating GTP molecules from GDP molecules:
\st
\be
\Delta_1 + \GDP \longrightarrow \GTP.
\ee
(ii) generating tubulin GTP from tubulin GDP:
\st
\be
\Delta_2+ \Tu_\GDP  ~\longrightarrow~ \Tu_\GTP ,
\ee
(iii) growth of an MT:
\st
\be
\Delta_3 + \MT_{N-1} + \Tu_\GTP ~\longrightarrow~ \MT_N,
\ee
(iv) shrinkage of an MT:
\st
\be
\MT_N  ~\longrightarrow~  \MT_{N-1} + \Tu_\GDP + \Delta_4.
\ee
Note that experimental studies determined the values of the free energies for these
reactions as:
$\Delta_1 \simeq 220$ meV, $\Delta_2 \simeq 160$ meV and
$\Delta_3 \simeq \Delta_4 \simeq 40$ meV \citep{Cap94}.  These free energies are clearly above
the thermal energy at room temperature ($kT\simeq 26$ meV) and
they are within a quantum mechanical energy range that corresponds to the creation of
one or a few chemical bounds.
Therefore, one may need to consider each chemical reaction as a quantum mechanics process.

In this paper, in order to simplify the problem we combine the above processes into two
fundamental reactions:\\
(i) growth of an MT by one dimer by adding of one tubulin layer in
an endothermic process:
\st
\be
\Delta + \MT_{N-1} + \Tu_\GTP ~\longrightarrow~ \MT_N,
\ee
(ii) shrinkage of an MT by one dimer due to the removal of one layer of $\Tu_\GDP$ dimer in
an exothermic process:
\st
\be
\MT_N  ~\longrightarrow~  \MT_{N-1} + \Tu_\GDP + \Delta,
\ee
where $\Delta$ is the energy of the reaction.
In order to derive a quantum mechanical description of mechanisms (i) and (ii), we first
need to introduce quantum states of MT, tubulin and heat bath:
\begin{itemize}
\item $|N\rangle$ is the state of a microtubule with $N$ dimers (both GTP and GDP
tubulins).
\item $|N_T\rangle$ is the state of a tubulin, $\Tu_\GTP$ or $\Tu_\GDP$.
\item $|\tilde{N}\rangle$ is the GTP hydrolysis energy state.
\end{itemize}
Then, the relevant second quantization operators would be:
\bea
\st\label{a_dagger}
&&a^\dagger=|N+1\rangle\langle N|,\\
\st
&&a=|N-1\rangle\langle N|,\\
\st
&&b^\dagger=|N_T+1\rangle\langle N_T|,\\
\st
&&b=|N_T-1\rangle\langle N_T|,\\
\st
&&d^\dagger=|\tilde{N}+1\rangle\langle \tilde{N}|,\\
\st\label{d}
&&d=|\tilde{N}-1\rangle\langle \tilde{N}|,
\eea
Here $b$/$b^\dagger$ and $d$/$d^\dagger$ are annihilation/creation operators
of tubulin and energy quanta, respectively.  The operators $a$/$a^\dagger$ are
lowering/raising the number of tubulin layers that constructed a MT.
Following \cite{TD01}, one can express the above processes using creation and
annihilation operators (\ref{a_dagger})-(\ref{d}):
\bea
\st\label{met1}
a^\dagger b ~d  &:& \Delta + \MT_{N-1} + \Tu_\GTP \longrightarrow  \MT_N \\
\st\label{met2}
d^\dagger ~ b^\dagger ~ a &:& \hspace{2.8cm}\MT_N  \longrightarrow
\MT_{N-1} +  \Tu_\GDP + \Delta
\eea
Operators (\ref{met1}) and (\ref{met2}) describe an MT's growth and shrinkage
by one layer, respectively.  Realistically, the polymerization or depolymerization
process may happen repeatedly before reversing the process.  This can be extended within
our model by constructing product operators, i.e. $(a^\dagger b ~d)^m$ and
$(d^\dagger ~ b^\dagger ~ a)^n$, where $m$ and $n$ are the number of growing or shrinking events in
a sequence, respectively.

\subsection{The Hamiltonian}
\noindent
Based on the mechanisms in (\ref{met1}) and (\ref{met2}), the Hamiltonian
for interacting microtubules with $\Tu_\GTP/\Tu_\GDP$ tubulins can be written as
\st
\bea\label{Ham}
H&=& \sum_\k \hbar\omega_\k a_\k^\dagger a_\k
+  \sum_\m \hbar\varpi_\m b_\m^\dagger b_\m
+  \sum_\l \hbar\sigma_\l~ d_\l^\dagger d_\l\no\\
&& \hspace{2cm} +\sum_{\k,\m} \hbar( \Delta_{\k,\m} ~a^\dagger_\k b_\m ~d_{\k-\m}
+   \Delta^*_{\k,\m} ~ d^\dagger_{\k-\m}  ~b^\dagger_\m ~ a_\k),
\eea
where $\omega$, $\varpi$, $\tilde{\Delta}$ and $\Delta$ are constants in units of energy.
However, an intermediate transition between a microtubule in a growing
phase and a microtubule in a shrinking phase must also be taken into account.
A growing/shrinking microtubule may change its state
quickly or after several steps to a depolymerizing/polymerizing state and then
may change back to polymerizing/depolymerizing state.
Experimentally, the transition of microtubules from the growing to the shrinking phase
is quantified by the catastrophe rate $f_{\rm cat}$ and the transition from the shrinking to
the growing phase is expressed by
the rescue rate $f_{\rm res}$ in which $f_{\rm res} < f_{\rm cat}$.   As we discussed earlier,
these transitions can be represented
by a combination of creation and annihilation operators as the $n^{\rm th}$ power of the reaction in
(\ref{met1}) and (\ref{met2}):
\st
\bea\label{Ham1}
H&=& \sum_\k \hbar\omega_\k a_\k^\dagger a_\k
+  \sum_\m \hbar\varpi_\m b_\m^\dagger b_\m
+  \sum_\l \hbar\sigma_\l~ d_\l^\dagger d_\l \no\\
&&\hspace{2cm}+ \sum_{n=1}^\infty \sum_{\tilde{\k}_n, \tilde{\m}_n, \tilde{\l}_{n-1}}
  \hbar[ \Delta_{\tilde{\k}_n \tilde{\m}_n \tilde{\l}_n} ~ c_{\tilde{\k}_n
  \tilde{\m}_n \tilde{\l}_n}
+   \Delta^*_{\tilde{\k}_n \tilde{\m}_n \tilde{\l}_n } ~
c^\dagger_{\tilde{\k}_n \tilde{\m}_n \tilde{\l}_n} ],
\eea
where
\st
\be\label{c-op}
c_{\tilde{\k}_n \tilde{\m}_n } =
(a^\dagger_{\k_1} b_{\m_1} ~d_{\l_1}) (a^\dagger_{\k_2} b_{\m_2} ~d_{\l_2})
\ldots
(a^\dagger_{\k_n} b_{\m_n} ~d_{\l_n }).
\ee
Here $\tilde{\k}_n=\{\k_1, \k_2, \ldots, \k_n\}$ is a collection of indices and
$\sum_{\tilde{\k}_n}=\sum_{\k_1} \sum_{\k_2} \ldots \sum_{\k_n}$.
We note that the momentum conservation for the last two terms in the
Hamiltonian (\ref{Ham1}) requires that
\st
\be\label{ln}
\l_n = \sum_{i=1}^n \k_i - \sum_{i=1}^n \m_i - \sum_{i=1}^{n-1} \l_i.
\ee
Therefore, the first $n-1$ of $\l$ will be free and summed in the Hamiltonian (\ref{Ham1}).

In Bose-Einstein statistics the creation and annihilation operators satisfy
\st
\be
[ q_\k, q_\m^\dagger]=\delta_{\k \m},~~{\rm and}~~
[q_\k^\dagger,q_\m^\dagger]=0=[q_\k,q_\m],
\ee
where $[A,B]=AB-BA$ is the Dirac commutator and $q=a,b,~{\rm and}~ d$ .  Since these
operators mutually commute,
the $c_{\tilde{\k}_n \tilde{\m}_n \tilde{\l}_n}$,  Eq. (\ref{c-op}), can be rewritten as
\st
\be\label{c-op1}
c_{\tilde{\k}_n \tilde{\m}_n \tilde{\l}_n} =
a^\dagger_{\k_1} a^\dagger_{\k_2}\ldots a^\dagger_{\k_n}  b_{\m_1} b_{\m_2} ~  \ldots
b_{\m_n} ~ d_{\l_1 } d_{\l_2} \ldots d_{\l_n} =  a^\dagger_{\tilde{\k}_n}
b_{\tilde{\m}_n} d_{\tilde{\l}_n},
\ee
where $\l_n$ is given by Eq. (\ref{ln}).

\section{Derivation of the equations of motion}

\noindent
The Heisenberg's equation of motion for a space- and time-dependent operator $q(\r,t)$
reads as
\st
\be
i\hbar \partial_t{q}(\r,t) = - [H,q(\r,t)],
\ee
where $H$ is the Hamiltonian.  Before finding equations
of motion, one needs
to calculate the commutation relation $[ q_\etas, q^\dagger_{\tilde{\k}_n}]$ that is
\st
\be\label{qq0}
[ q_\etas, q^\dagger_{\tilde{\k}_n}]=[q_\etas, q^\dagger_{\k_1} \ldots
q^\dagger_{\k_n}] =
 \delta_{\etas,  \k_1}~ q^\dagger_{\k_2} q^\dagger_{\k_3} \ldots q^\dagger_{\k_n  }
+  \delta_{\etas,  \k_2}~ q^\dagger_{\k_1} q^\dagger_{\k_3} \ldots q^\dagger_{\k_n  }
+ \ldots +
 \delta_{\etas,  \k_n}~  q^\dagger_{\k_1} q^\dagger_{\k_2} \ldots q^\dagger_{\k_{n-1}}.
\ee
Since all $\k_1, \k_2, \ldots, \k_n$ are dummy indices one can write Eq. (\ref{qq0}) as
\st
\be\label{qq}
[ q_\etas, q^\dagger_{\tilde{\k}_n}]= n~\delta_{\etas, \k_n} ~q^\dagger_{\tilde{\k}_{n-1}},
\ee
where $\k_n$ is chosen for simplicity.
Using Eq. (\ref{qq}) we can find the commutation relations between
$a_\etas$ and $b_\etas$
operators with $c_{\tilde{\k}_n \tilde{\m}_n \tilde{\l}_n}$ and
$c_{\tilde{\k}_n \tilde{\m}_n \tilde{\l}_n}^\dagger$ operators as
\st
\bea
&&[ a_\etas, c_{\tilde{\k}_n \tilde{\m}_n \tilde{\l}_n}  ]
= [ a_\etas, a^\dagger_{\tilde{\k}_n} b_{\tilde{\m}_n} d_{\tilde{\l}_n } ]
= n~\delta_{ \etas, \k_n  } a^\dagger_{\tilde{\k}_{n-1}} b_{\tilde{\m}_n} d_{\tilde{\l}_n}, \\
\st
&&[ b_\etas, c^\dagger_{\tilde{\k}_n \tilde{\m}_n \tilde{\l}_n } ]
= [ b_\etas,   d^\dagger_{\tilde{\l}_n } b^\dagger_{\tilde{\m}_n}
a_{\tilde{\k}_n} ]
= n~\delta_{ \etas, \m_n } d^\dagger_{\tilde{\l}_n }
b^\dagger_{\tilde{\m}_{n-1}} a_{\tilde{\k}_n}.
\eea
However, the commutation relation between $d_\etas$ operator and
$c_{\tilde{\k}_n \tilde{\m}_n \tilde{\l}_n}^\dagger$
will be
\st
\be
[ d_\etas, c^\dagger_{\tilde{\k}_n \tilde{\m}_n \tilde{\l}_n } ]
= [ d_\etas,   d^\dagger_{\tilde{\l}_n } b^\dagger_{\tilde{\m}_n}
a_{\tilde{\k}_n} ]
= \left( (n-1)~ \delta_{ \etas, \l_{n-1} } d^\dagger_{\tilde{\l}_{n-2} }  d^\dagger_{\l_n}
+ ~ \delta_{ \etas, \l_{n} } d^\dagger_{\tilde{\l}_{n-1} }  \right)
 b^\dagger_{\tilde{\m}_n}  a_{\tilde{\k}_n},
\ee
where $\l_n$ is given by Eq. (\ref{ln}).
Therefore, the equation of motion for
$a_\etas$, $b_\etas$ and $d_\etas$ operators can be derived from Hamiltonian (\ref{Ham1}) as
\bea
\st\label{eq1}
&&i\partial_t{a}_\etas = \omega_\etas a_\etas
          +  \sum_n  \sum_{\tilde{\k}_{n-1} \tilde{\m}_n \tilde{\l}_{n-1}}
  n~ \Delta_{\etas \tilde{\k}_{n-1} \tilde{\m}_n \tilde{\l}_{n-1}}
  a^\dagger_{\tilde{\k}_{n-1}} b_{\tilde{\m}_{n}}  d_{\tilde{\l}_{n-1}}
d_{\etas+\sum_{i=1}^{n-1} (\k_i-\l_i) - \sum_{i=1}^n \m_i  }
,~~~~\\
\st\label{eq2}
&&i \partial_t{b}_\etas = \varpi_\etas b_\etas +
\sum_n  \sum_{\tilde{\k}_{n} \tilde{\m}_{n-1} \tilde{\l}_{n-1}}
  n~ \Delta_{\etas \tilde{\k}_{n} \tilde{\m}_{n-1} \tilde{\l}_{n-1}}
 ~ d^\dagger_{\tilde{\l}_{n-1} }
d^\dagger_{{\sum_{i=1}^{n} \k_i - \etas- \sum_{i=1}^{n-1}( \m_i + \l_i)} }
b^\dagger_{\tilde{\m}_{n-1}}
 a_{\tilde{\k}_{n}}
,~~~~\\
\st\label{eq3}
&&i \partial_t{d}_\etas = \sigma_\etas d_\etas
+  \sum_n  \sum_{\tilde{\k}_n \tilde{\m}_{n} \tilde{\l}_{n-2}}
(n-1)~\Delta_{\etas \tilde{\k}_n \tilde{\m}_{n}\tilde{\l}_{n-2} }
d^\dagger_{\tilde{\l}_{n-2}} d^\dagger_{{\sum_{i=1}^{n} (\k_i-\m_i) -
\etas- \sum_{i=1}^{n-2} \l_i} }
 ~b^\dagger_{\tilde{\m}_{n}}
a_{\tilde{\k}_{n}}\no\\
&&\hspace{2.5cm}
+  \sum_n  \sum_{\tilde{\k}_n \tilde{\m}_{n} \tilde{\l}_{n-1}}
\delta_{\etas, {{\sum_{i=1}^{n} (\k_i-\m_i) - \sum_{i=1}^{n-1} \l_i} }}
 ~\Delta_{\tilde{\k}_n \tilde{\m}_{n}\tilde{\l}_{n-1} }
d^\dagger_{\tilde{\l}_{n-1}}
 ~b^\dagger_{\tilde{\m}_{n}}
a_{\tilde{\k}_{n}}.
\eea
Equations (\ref{eq1})-(\ref{eq3}) describe the dynamics of an MT in a quantum manner.
Since MTs are overall classical objects, we need to ensemble average over all possible
states to obtain effective dynamical equations.

\section{Classical equations of motion}
\noindent
Fourier transforming of $a_\etas$, $b_\etas$ and $d_\etas$ operators
over all states, one can find
\bea
\st
&&\psi(\r,t) = \Omega^{-1/2} \sum_\etas \exp(-i \eeta\cdot\r) a_\etas (t),\\
\st
&&\chi(\r,t) = \Omega^{-1/2} \sum_\etas \exp(-i \eeta\cdot\r) b_\etas (t),\\
\st
&&\phi(\r,t) = \Omega^{-1/2} \sum_\etas \exp(-i \eeta\cdot\r) d_\etas (t),
\eea
where $\Omega$ is the volume over which the members of the plane wave basis are
normalized \citep{TD89a,DT95}.   Here $\psi(\r,t)$, $\chi(\r,t)$, and $\phi(\r,t)$
are corresponding field operators for the quantum operators $a_\etas$, $b_\etas$ and
$d_\etas$, respectively.   The derivation of the equation of motion for the field
operators are given in Appendix.
The final form of the equations of motion is found to be
\st
\bea\label{eq1_e}
&& \partial_t{\psi} = A_0 \psi + \A_1\cdot \nab\psi
+ D_0 \nabla^2 \psi + \sum_{n=2}^\infty ~ (A_2^{(n)}\psip) ~\psip^{n-2}\chi^n\phi^n, \\
\st\label{eq2_d}
&& \partial_t{\chi} = B_0 \chi + D_1\nabla^2 \chi
+ \sum_{n=1}^\infty  (B_1^{(n)} \psi) \psi^{n-1}\chip^{n-1}\phip^n , \\
\st\label{eq3_d}
&&\partial_t{\phi} = C_0 \phi + D_2\nabla^2 \phi
+ \sum_{n=1}^\infty (C_1^{(n)} \psi)\psi^{n-1}\chip^n\phip^{n-1},
\eea
where $n$ represents the degree of nonlinearity and  $A_i$, $B_i$, $C_i$ and $D_i$ are constants and given in Appendix.
We obtain the general equations of motion for the system in terms of coupled nonlinear partial
differential equations (PDE's) that describe the MT field, the tubulin field and GTP
field, respectively.

\noindent
In this paper we are primarily interested in the dynamics of MTs.
Following Eq. (\ref{eq1_e}), the dynamical equations for
growing and shrinking states of an MT up to $n=3$ can be written as
\bea
\st\label{eqMT_s}
&& \partial_t{\psi} = A_0 \psi + \A_1\cdot \nab\psi
+ D_0\nabla^2 \psi + (A_2^{(2)} \psip )~\chi^2\phi^2 + (A_2^{(3)}\psip )~\psip\chi^3\phi^3.
\eea
Here $\psip$/$\psi$ represent the growing/shrinking state of the MT.
Furthermore, the dynamics of the tubulin, $\chi$, and energy of the system, $\phi$, are
also determined by
\bea
\st\label{eq_chi}
\partial_t{\chi} &=& B_0 \chi + D_1\nabla^2 \chi + (B_1^{(1)}\psi ) \phip , \\
\st\label{eq_phi}
\partial_t{\phi} &=& C_0 \phi + D_2\nabla^2 \phi + (C_1^{(1)}\psi ) \chip,
\eea
where, for simplicity, we just keep the $n=1$ term.

It is clear that the system of equations (\ref{eqMT_s})-(\ref{eq_phi})
is very similar to the phenomenological system of equations
(\ref{grow})-(\ref{c_D}) which has been extensively studied in the
nonlinear physics literature.   A vast array of mathematical methods of finding
their solutions can be found in the monograph by \cite{Dix97}.
Among them one can expect to find localized (solitonic) and extended (traveling wave) solutions.
The latter ones may have the meaning of coherent oscillations observed experimentally
for high tubulin concentrations by \cite{Man89}.

\section{Discussion}
\noindent
In our model, the basic structural unit is the tubulin dimer. Each dimer exists in a quantum
mechanical state characterized by several variables even in our simplified approach.
Each microstate of a tubulin dimer  is sensitive to the states of its neighbors.
Tubulin dimers have both discrete degrees of freedom (distribution of charge) and
continuous degrees of freedom (orientation). A model that focuses on the discrete
will be an array of coupled binary switches \citep{Ras90,Cam01},
while a model that focuses on the continuous will probably be an array of coupled
oscillators \citep{Sam92,BT97}.  In the present paper
we have focused on tubulin binding and GTP hydrolysis as the key processes determining
the states of microtubules. These are also the degrees of freedom that are most easily
accessible to experimental determination.  In this paper we have shown how a quantum
mechanical description of the energy binding reactions taking place during MT polymerization
can be led to nonlinear field dynamics with very rich behavior that includes both localized
energy transfer and oscillatory solutions.

We have demonstrated here that the
assembly process can be described using quantum mechanical principles applied to biochemical
reactions.   This can be subsequently transformed into a highly nonlinear
semi-classical dynamics problem.  The gross features of MT dynamics satisfy
classical field equations in a coarse-grained picture.  Individual chemical
reactions involving the constituent molecules still retain their quantum
character.  The method of coherent structures allows for a simultaneous
classical representation of the field variables and a quantum approach to their
fluctuations.   Here, the overall MT structure (and their ensembles) can be
viewed as a virtual classical object in (3+1) dimensional space.   However,
at the fundamental level of its constituent biomolecules, it is quantized as
are true chemical reactions involving its assembly or disassembly.   Whether
this process can be implicated in nonlinear computation or information processing by
neurons is an open question.  The main problem of quantum
computation is decoherence: the loss of entanglement from within the quantum
computer into its environment. If there is no entanglement left, there is no
quantum parallelism, only a stochastic process with no advantage over classical
computation \citep{AB96}.  The best-known model of quantum computation
in the microtubule \cite{HP96} is a quantized cellular automaton model,
with the additional postulate that state reduction occurs spontaneously \citep{Pen94}.
Its decoherence timescale has been estimated at $10^{-5}$ to $10^{-4}$ seconds without
shielding of the microtubule, and $10^{-2}$ to $10^{-1}$ seconds with shielding
by an actin gel
present in the cell \citep{Hag00}. However, the original classical model has
been criticized as unrealistic \citep{BT97}, and the proposed alternative
(a continuum model) decoheres much more rapidly \citep{Teg00}, suggesting that it
can only function classically.  In our approach the route taken is opposite since we
started with individual tubulin quantum microstates to arrive at classical,
nonlinear but coherent (and stable) macro-states of a microtubule.

Interestingly, using a simplified model of the dimer as a double potential well, the conductivity
of the microtubule was recently calculated \citep{BT01}. For a micron-long
microtubule, the predicted value falls into the `good intrinsic semiconductor' regime,
and even reaches the semi-metallic regime at high electron concentrations.  The length of
a microtubule is directly proportional to its resistance via Ohm's law hence there exists
a direct link between conductive properties of microtubules and their length which is
seen in our model as the average of a mesoscopic state (probability density wave).

The search is still on for a realistic model of  quantum computation in the microtubule,
one that is grounded in the 1998 atomic structural data. If such a model is found, many
more questions will be raised:

\emph{How do separate microtubules become entangled}?
It seems unlikely that quantum coherence
would be limited to individual microtubules. One possibility is that electrons or
quasiparticles tunneling through MAPs cause associated microtubules within the same
cell to become entangled. A less likely possibility is that electromagnetic fields
generated by individual microtubules do the trick. Looking beyond the single cell,
it has been proposed that electrons could even tunnel from one cell to the next
through a gap junction, a transient fusion of the membranes of neighboring cells
\citep{WH01}.

In our model, an ensemble of MT's can become a higher
level coherent structure if the tubulin density is sufficiently high to result
in significant correlations between individual MT's and their interactions leading
to the experimentally observed synchronization of MT assembly and coherent oscillations
in the assembled tubulin mass \citep{Man89}.

Finally, contact must eventually be made with experiment. A starting point would be
to learn more about electron motion in microtubules. \cite{Bec75} demonstrated
the existence of fluorescent resonant energy transfer between aromatics in adjacent
tubulins, and between microtubules and membranes. Such exchanges might serve to power
the motion of electrons through the aromatic lattice without dissipation of energy.
A new generation of such experiments, under varied conditions of pH, MAP attachment,
and so forth, could be very helpful both in building and testing models.

On the computational side, it has been suggested that the principal output of
microtubules takes the form of highly symmetrical MAP attachment patterns
\citep{Sam92, WH01} which determine subsequent
cytoskeletal growth and behavior. In this case, progress in understanding
microtubular computation will be measured by the ability to interpret and predict
these outputs.  If microtubules are indeed information processors, it seems likely
that a long period of trial and error will be necessary before we truly learn how
they work.

\begin{acknowledgments}
This research was supported in part by the Natural Sciences and
Engineering Research Council of Canada (NSERC) and the Canadian Space Agency (CSA).
Insightful discussions with Prof. S. R. Hameroff and Dr. J. M. Dixon are gratefully
acknowledged.
\end{acknowledgments}

\setcounter{sub}{0}
\setcounter{subeqn}{0}
\renewcommand{\theequation}{A.\thesub\thesubeqn}

\section*{Appendix A. Derivation of equation of motion for the field operators}
\noindent
Multiplying both sides of Eq. (\ref{eq1}) by $\exp(-i\eeta\cdot\r)$, dividing by
$\Omega^{1/2}$ and
summing over $\eeta$, one finds
\st
\bea\label{eq1_a}
&&i\partial_t\psi_ = \Omega^{-1/2}\ms \sum_\etas \omega_\etas \exp(-i\eeta\cdot\r)
a_\etas\right.\no\\
&&\left. \hspace{0cm}  + \sum_n  \sum_\etas \sum_{\tilde{\k}_{n}
\tilde{\m}_n \tilde{\l}_{n-1}}
n~ \Delta_{\etas \tilde{\k}_{n-1} \tilde{\m}_n \tilde{\l}_{n-1}} \exp(-i\eeta\cdot\r)~
  a^\dagger_{\tilde{\k}_{n-1}} b_{\tilde{\m}_{n}}  d_{\tilde{\l}_{n-1}}
d_{\etas+\sum_{j=1}^{n-1} (\k_j-\l_j) - \sum_{j=1}^n \m_j  }
 \rs.\no\\
\eea
Changing $\eeta \rightarrow
\eeta - \sum_{j=1}^{n-1} (\k_j-\l_j) + \sum_{j=1}^n \m_j  $ in the
second term of Eq. (\ref{eq1_a}), one  finds
\st
\bea\label{eq1_b}
&&i \partial_t\psi_ = \Omega^{-1/2}\ms \sum_\etas \omega_\etas \exp(-i\eeta\cdot\r)
a_\etas\right.\no\\
&& \hspace{3cm}  +
\sum_n  \sum_\etas \sum_{\tilde{\k}_{n} \tilde{\m}_n \tilde{\l}_{n-1}}
n~ \Delta_{\etas-\xis~ \tilde{\k}_{n-1} \tilde{\m}_n \tilde{\l}_{n-1}}
e^{-i\etas\cdot\r + i \sum_{j=1}^{n-1} (\k_j-\l_j)\cdot \r - i \sum_{j=1}^n \m_j\cdot\r}\no\\
&&\left. \hspace{5cm} \times ~
  a^\dagger_{\tilde{\k}_{n-1}} b_{\tilde{\m}_{n}}  d_{\tilde{\l}_{n-1}}
d_\etas
\rs,
\eea
or
\st
\bea\label{eq1_c}
&&i \partial_t\psi_ = \Omega^{-1/2}\ms \sum_\etas \omega_\etas \exp(-i\eeta\cdot\r)
a_\etas + ~\sum_n \sum_\etas \sum_{\tilde{\k}_{n} \tilde{\m}_n \tilde{\l}_{n-1}}
n~ \Delta_{\etas-\xis~ \tilde{\k}_{n-1} \tilde{\m}_n \tilde{\l}_{n-1}} \right. \no\\
&& \hspace{4cm}\left. \times
~e^{-i\etas\cdot\r}~ d_\etas~
e^{i \tilde{\k}_{n-1}\cdot \r}~ a^\dagger_{\tilde{\k}_{n-1}} ~
e^{-i \tilde{\m}_{n}\cdot\r} ~b_{\tilde{\m}_{n}}  ~
e^{-i \tilde{\l}_{n-1}\cdot \r}~ d_{\tilde{\l}_{n-1}}~ \frac{}{}
\rs,
\eea
where $\xxi =  \sum_{j=1}^{n-1} (\k_j-\l_j) - \sum_{j=1}^n \m_j$.
Here, for example,
$\exp(-i\tilde{\k}_n\cdot \r) = \exp(-i\k_1\cdot \r) \exp(-i\k_2\cdot \r) \ldots
\exp(-i\k_n\cdot \r) = \exp(-i\sum_{j=1}^n \k_j\cdot \r)$.
Our goal is now to rewrite Eq. (\ref{eq1_c}) in terms of field operators,
$\psi,~ \chi,~ \phi$,
and their derivatives.  This can be done in a straightforward manner provided the
dispersion matrix elements $\omega_\etas$ and
$ \Delta_{\etas-\xis~ \tilde{\k}_{n-1} \tilde{\m}_{n} \tilde{\l}_{n-1}}$
which are generally
function $\eeta$, $\k_i$, $\m_i$ and $\l_i$
($1\leq i \leq n$) are known.  Unfortunately, such information is very model
dependent.  Therefore, the simplest
way that also keeps the generality of the problem is to Taylor expand these
matrix elements about some point
($\eeta_0,\k_{0i},\m_{0i}, \l_{0i}$) in the space spanned by $\eeta$,
$\k_i$, $\m_i$ and $\l_i$ \citep{TD89a,DT95}.

Expanding $\omega_\etas$ to all orders, one finds
\st
\be\label{omega}
\omega_\etas = \omega_0 + \sum_{s=1}^\infty
[(\eeta-\eeta_0)\cdot \nab_\etas]^s\omega_0/s!\;,
\ee
where $\omega_0 =\omega_{\etas_0}$.  Furthermore, for any function
$ f(\eeta, \tilde{\k}_n, \tilde{\m}_n, \tilde{\l}_n)= \Delta_{\etas \tilde{\k}_{n} \tilde{\m}_{n} \tilde{\l}_{n}}$
we can write
\st
\bea\label{f}
&&f(\eeta, \tilde{\k}_n, \tilde{\m}_n, \tilde{\l}_n) = f_0
+ (\eeta-\eeta_0)\cdot\nabla_\etas f|_0
+ \sum_{j=1}^n(\k_j-\k_{0j})\cdot\nab_{\k_j} f|_0
\no\\
&&\hspace{3.5cm}
+ \sum_{j=1}^n(\m_j-\m_{0j})\cdot\nabla_{\m_j} f|_0
+ \sum_{j=1}^n(\l_j-\l_{0j})\cdot\nab_{\l_j} f|_0 + \ldots,
\no\\
&& \hspace{.2cm}+ \sum_{p,q,r=1}^n \sum_{s=2}^\infty \sum_{u=0}^s
\sum_{v=0}^{u} \sum_{w=0}^{s-u-v}
 {^s}C_u~~ {^{u}}C_v ~~ {^{s-u-v}}C_w /s! ~~ ~\no\\
&& \hspace{1cm} \times ~
 [(\eeta-\eeta_0)\cdot \nab_\etas]^u~
[(\k_p-\k_{0p})\cdot \nab_{\k_p}]^v ~
[(\m_q-\m_{0q})\cdot \nab_{\m_q}]^w ~[(\l_r-\l_{0r})\cdot
\nab_{\l_r}]^{s-u-v-w}f|_0,\no\\
\eea
where
\st
\be
f_0 = \sum_{p, q, r=1}^n f(\eeta_0,\k_{0p}, \m_{0q}, \l_{0r}),
\ee
where $^sC_r$ are binomial coefficients.
Here, for example, $\nab_{\m} f$ means $\hat{i} \partial_{m_{x}} f  + \hat{j}
\partial_{m_{y}} f + \hat{k} \partial_{m_{z}} f$ where $\hat{i}, \hat{j}$
and $\hat{k}$ are unit vectors in the $m_x, m_y$ and $m_z$ directions,
respectively, and $\nab_\m f|_0$ is the value of the gradient at point
($\eeta_0, \k_0, \m_0, \l_0$).

Using Eqs. (\ref{omega}) and (\ref{f}), Eq. (\ref{eq1_c}) can be written as
\st
\bea\label{eq1_d}
&&i \partial_t\psi = \lambda_0(\omega) \psi + i\llambda_1(\omega)\cdot \nab\psi
- {1\over 2} \sum_{i,j} [\llambda_2(\omega)]_{ij}\partial_{x_i x_j}^2 \psi
+ \sum_n n \Omega^{(3n-1)/2} \Lambda_{1}^{(n)} \psip^{n-1} \chi^n \phi^{n}\no\\
&& + \sum_n n \Omega^{(3n-1)/2} \left( \psip^{n-1}\chi^n\phi^{n-1}\nab_\etas f|_0
\cdot \nab\phi
+ \sum_{j=1}^{n-1} \nab_{\k_j}f|_0 \cdot \nab\psip ~\psip^{n-2}\chi^n\phi^n
 \right.\no\\
&&\hspace{2cm}+\left. \sum_{j=1}^{n} \psip^{n-1}\nab_{\m_j}f|_0 \cdot \nab\chi
~\chi^{n-1}\phi^n
+ \sum_{j=1}^{n-1}\psip^{n-1}\chi^n \nab_{\l_j}f|_0 \cdot \nab\phi~ \phi^{n-1} \right),
\eea
where
\st
\bea
&&\lambda_0(\omega) = \omega_0 - \eeta_0\cdot\nab_\eta\omega|_0 + (1/2)\sum_{i,j}
  \eta_{0i}\eta_{0j}
\partial^2_{\eta_i\eta_j}\omega|_0~, \\
\st
&&[\llambda_1(\omega)]_i = -\sum_j\eta_{0j}\partial^2_{\eta_i\eta_j} \omega|_0
+ \partial_{\eta_i} \omega|_0~,\\
\st
&&[\llambda_2(\omega)]_{ij} = \partial^2_{\eta_i\eta_j} \omega|_0,\\
\st
&& \Lambda_{1}^{(n)} = f_0 -\eeta_0\cdot\nab_\eta f|_0 - \sum_{j=1}^{n-1}
\k_{0j}\cdot\nab_{k_j} f|_0
- \sum_{j=1}^{n} \m_{0j}\cdot\nab_{m_j}f|_0  - \sum_{j=1}^{n-1}
 \l_{0j}\cdot\nab_{l_j} f|_0.~~~~~~~~
\eea
Similarly, using Eqs. (\ref{eq2}) and (\ref{eq3}), one can write equations of
 motion for $\chi$ and $\phi$
as
\st
\bea\label{eq2_b}
&&i \partial_t\chi = \lambda_0(\varpi) \chi + i\llambda_1(\varpi)\cdot \nab\chi
- {1\over 2} \sum_{i,j} [\llambda_2(\varpi)]_{ij}\partial_{x_i x_j}^2 \chi
+ \sum_n n \Omega^{(3n-1)/2} \Lambda_2^{(n)} \psi^{n} \chip^{n-1} \phip^{n}\no\\
&& + \sum_n n \Omega^{(3n-1)/2} \left( \psi^{n}\chip^{n-1}\phip^{n-1}\nab_\etas
f|_0 \cdot \nab\phip
+ \sum_{j=1}^{n} \nab_{\k_j}f|_0 \cdot \nab\psi ~\psi^{n-1}\chip^n\phip^n
 \right.\no\\
&&\hspace{2cm}+\left. \sum_{j=2}^{n-1} \psi^{n}\nab_{\m_j}f|_0 \cdot \nab\chip
~\chi^{n-2}\phi^n
+ \sum_{j=1}^{n-1}\psi^{n}\chip^{n-1} \nab_{\l_j}f|_0 \cdot \nab\phip~ \phip^{n-1} \right),\\
\st\label{eq3_b}
&&i\partial_t\phi = \lambda_0(\sigma) \phi + i\llambda_1(\sigma)\cdot \nab\phi
- {1\over 2} \sum_{i,j} [\llambda_2(\sigma)]_{ij}\partial_{x_i x_j}^2 \phi
+ \sum_n n \Omega^{(3n-1)/2} \Lambda_3^{(n)} \psi^{n} \chip^n \phip^{n-1}\no\\
&& + \sum_n \Omega^{(3n-1)/2} \left( (n-1) \psi^{n}\chip^n\phip^{n-2}
\nab_\etas f|_0 \cdot \nab\phip
+  n  \sum_{j=1}^{n} \nab_{\k_j}f|_0 \cdot \nab\psi ~\psi^{n-1}\chip^n\phi^{n-1}  \right.\no\\
&&\hspace{2cm}+\left. n\sum_{j=1}^{n} \psi^{n}\nab_{\m_j}f|_0 \cdot \nab\chip
 ~\chip^{n-1}\phip^{n-1}
+ n\sum_{j=1}^{n-1}\psi^{n}\chip^n \nab_{\l_j}f|_0 \cdot \nab\phip~ \phip^{n-2} \right),\no\\
\eea
where
\st
\bea
&& \Lambda_{2}^{(n)} = f_0 -\eeta_0\cdot\nab_\eta f|_0 - \sum_{j=1}^{n}
\k_{0j}\cdot\nab_{k_j} f|_0
- \sum_{j=1}^{n-1} \m_{0j}\cdot\nab_{m_j}f|_0  - \sum_{j=1}^{n-1}
 \l_{0j}\cdot\nab_{l_j} f|_0,~~~~~~~~~\\
\st
&& \Lambda_{3}^{(n)} = f_0 -\eeta_0\cdot\nab_\eta f|_0 - \sum_{j=1}^{n}
\k_{0j}\cdot\nab_{k_j} f|_0
- \sum_{j=1}^{n} \m_{0j}\cdot\nab_{m_j}f|_0  - \sum_{j=1}^{n-1}
 \l_{0j}\cdot\nab_{l_j} f|_0.~~~~~~~~~
\eea

Simplifying the equations of motion as
\st
\bea\label{eq1_eA}
&& \partial_t{\psi} = A_0 \psi + \A_1\cdot \nab\psi
+ D_0 \nabla^2 \psi+ \sum_n ~( A_2^{(n)} \psip)~\psip^{n-2}\chi^n\phi^n, \\
\st\label{eq2_dA}
&& \partial_t{\chi} = B_0 \chi + D_1\nabla^2 \chi
+ \sum_n ( B_1^{(n)} \psi) \psi^{n-1}\chip^{n-1}\phip^n ,\no \\
&&\\
\st\label{eq3_dA}
&&\partial_t{\phi} = C_0 \phi + D_2\nabla^2 \phi
+ \sum_n  ( C_1^{(n)} \psi )  \psi^{n-1}\chip^n\phip^{n-1},
\eea
where
\st
\bea
&&A_0 = -i\lambda_0(\omega),~~\A_1=\llambda_1(\omega),~~
A_2^{(n)}\psip= -i n \Omega^{3n-1\over 2}(\Lambda_1^{(n)} +  \sum_{j=1}^{n-1} \nab_{\k_j}f|_0 \cdot \nab)\psip ,~~~~~~\\
\st
&&B_0 = -i\lambda_0(\varpi),~~
B_1^{(n)}\psi= - i n \Omega^{3n-1\over 2}( \Lambda_2^{(n)} + \sum_{j=1}^{n} \nab_{\k_j}f|_0 \cdot \nab)\psi  ,\\
\st
&&C_0 = -i\lambda_0(\sigma),~~
C_1^{(n)}\psi= - i n \Omega^{3n-1\over 2} ( \Lambda_3^{(n)} + \sum_{j=1}^{n} \nab_{\k_j}f|_0 \cdot \nab)\psi,\\
\st
&&D_0=2i\lambda_2(\omega),~~D_1=2i\lambda_2(\varpi),~~D_2=2i\lambda_i(\sigma),~~.
\eea



%
\def\pra{{Phys.~Rev.~A}}        
\def\prb{{Phys.~Rev.~B}}        
\def\prc{{Phys.~Rev.~C\ }}        
\def\prd{{Phys.~Rev.~D\ }}        
\def\pre{{Phys.~Rev.~E}}        
\def\prl{{Phys.~Rev.~Lett.\ }}    
\def\nat{{Nature\ }}              
\def\iaucirc{{IAU~Circ. No.}}       
\def\nphysa{{Nucl.~Phys.~A}}   
\def\nphysb{{Nucl.~Phys.~B\ }}   
\def\physrep{{Phys.~Rep.}}   
\def\physscr{{Phys.~Scr}}   

----------------------------------------------------------------------------------------

\end{document}